\documentclass[aps,preprint]{revtex4}

\usepackage{amsfonts}
\usepackage{amsmath}
\usepackage{amssymb}
\usepackage{subfigure}
\usepackage{graphicx}
\usepackage{epstopdf}
\usepackage{epsfig}
\usepackage{float}
\usepackage{color}
\usepackage{ulem}
\usepackage{cancel}
\usepackage{mathrsfs}
\usepackage[colorlinks,linkcolor=blue,citecolor=blue,urlcolor=blue]{hyperref}
\usepackage{url}
\usepackage{cleveref}
\usepackage{ulem}

\begin{document}

\preprint{CTP-SCU/2022011}
\title{Phase transitions and thermodynamic geometry of a Kerr-Newman black hole in a cavity}
\author{Yuchen Huang$^{a,b}$}
\email{ychuang@mail.ustc.edu.cn}
\author{Jun Tao$^{b}$}
\email{taojun@scu.edu.cn}
\author{Peng Wang$^{b}$}
\email{pengw@scu.edu.cn}
\author{Shuxuan Ying$^{c}$}
\email{ysxuan@cqu.edu.cn}

\affiliation{$^{a}$Center for Theoretical Physics, College of Physics, Sichuan University, Chengdu, 610065, China}
\affiliation{$^{b}$Department of Astronomy, School of Physical Sciences, University of Science and Technology of China, Hefei, 230026, China}
\affiliation{$^{c}$Department of Physics, Chongqing University, Chongqing, 401331, China}

\begin{abstract}

Being placed in a cavity is an effective way of reaching thermodynamic equilibrium for black holes. We investigate a Kerr-Newman black hole in a cavity as well as compare it with two reduced cases, i.e., a RN black hole in a cavity and a Kerr black hole in a cavity. We derive the quasi-local energy from the Hamiltonian, and construct the first law of thermodynamics accordingly. In a canonical ensemble, these black holes could undergo a van der Waals-like phase transition, which is very similar to that in AdS space. We further investigate the black holes' thermodynamic geometry, which is a powerful tool to diagnose microscopic interactions of a thermodynamic system. Our results show that in a cavity, although phase structures of these black holes are similar, their thermodynamic geometry show strong dissimilarities, implying that the microstructure of a black hole is sensitive to its states.

\end{abstract}

\maketitle

\section{Introduction}

How to define the quasi-local energy of a gravitational field in a spatially bounded cavity is an important topic in general relativity. For example, developing a tool to calculate the quasi-local energy could help us better understand the theory of gravitation itself. Moreover, correct thermodynamic quantities of black holes are probably at the quasi-local level. For the simplest non-trivial black hole solution in general relativity, the Schwarzschild black hole, whose thermodynamics in a cavity was first investigated with the help of the Euclidean action \cite{York:1986it}, where a phase transition that occurs between thermal flat spacetime and the black hole was found to be very similar to that of the Hawking-Page in AdS space \cite{Hawking:1982dh}. The thermodynamic quantities are consistent with the results from the later proposed Brown-York definition of quasi-local quantities \cite{Brown:1992br}. By means of such a definition, thermodynamic quantities for some simple gravitational systems enclosed by a cavity were obtained \cite{Brown:1994gs}. Afterwards, the thermodynamics and phase transitions of a RN black hole in a cavity were studied in a grand canonical ensemble \cite{Braden:1990hw} and a canonical ensemble \cite{Carlip:2003ne,Lundgren:2006kt}, which also exhibit some similarities to the black hole in asymptotic AdS space.

Recently, some works have been focusing on thermodynamics of various black holes in a cavity, which devoted to comparing phase transitions of black holes under different boundary conditions \cite{Wang:2019kxp,Liang:2019dni,Wang:2019urm,Huang:2021iyf,Yao:2021zid,Huang:2021eby}, i.e., AdS space and the cavity. A main conclusion is that the thermodynamics and phase transitions of black holes under the two different boundary conditions show close resemblances but several dissimilarities in details. Furthermore, the thermodynamics of a black hole in a cavity was discussed in an extended phase space, where volume of the cavity was treated as thermodynamic volume \cite{Wang:2020hjw}. On the whole, most discussions about the thermodynamics of black holes in a cavity especially phase transitions, were based on static black holes. Although a rotating BTZ black hole was studied in \cite{Huang:2021eby}, it did not show rich phase structures due to its non-negative heat capacity. Therefore, one naturally generalizes the study to a four-dimensional black hole with an angular momentum.

On the other hand, although black holes can undergo phase transitions, an complete quantum gravity theory that can explain them microscopically still remains absent. Nonetheless, there are many attempts trying to give microscopic interpretations of black hole thermodynamics. The thermodynamic geometry method, which sheds some fresh insights into equilibrium thermodynamics, has been proved to be a useful tool to probe microscopic structures of black holes. The idea of using the Riemannian geometry to study equilibrium thermodynamics was proposed by Weinhold \cite{1975JChPh..63.2479W}, who introduced a curvature scalar of a metric defined by the derivative of the internal energy with respect to the entropy and other extensive variables. Since this geometry seems meaningless in physics, Ruppeiner further developed the theory and introduced a new metric, but defined by the derivative of the entropy with respect to other extensive variables \cite{Ruppeiner:1995zz}. In this definition, the line element measures the distance between two neighboring fluctuation
states in equilibrium thermodynamic. The Riemann curvature scalar based on this metric is also called the Ruppeiner invariant. After computing the Ruppeiner invariant in various thermodynamic systems, such as ideal gas \cite{PhysRevA.20.1608,PhysRevA.41.2200}, ideal paramagnet \cite{PhysRevA.39.6515}, ideal quantum gas \cite{1990JPhA...23..467J}, one-dimensional Ising model \cite{PhysRevA.24.488}, and so forth, one came to an empirical conclusion that the positive/negative curvature scalar is caused by the repulsive/attractive interaction. This intriguing phenomenon was then generalized to the study of thermodynamics of black holes \cite{Ferrara:1997tw}. In the twenty years thereafter applications of the thermodynamic geometry in black holes attracted many theoretical interests \cite{Cai:1998ep,Aman:2003ug,Shen:2005nu,Aman:2005xk,Sarkar:2006tg,Quevedo:2008xn,Ruppeiner:2008kd,Banerjee:2010bx,Banerjee:2010da,Niu:2011tb,Zhang:2015ova,Wei:2015iwa,Wei:2019yvs,Wei:2019uqg,Wei:2020poh,Wei:2021lmo}. Nonetheless, the thermodynamic geometry of black holes in a cavity did not receive much attentions, which might help us to probe microstructures of black holes in a new boundary condition. Although some researches have made initial progress by comparing it in AdS space and a cavity \cite{Wang:2019cax,Wang:2021llu}, we point out that these studies are somewhat incomplete since they only considered the simply case of a RN black hole.

To deepen our understanding of microstructures of black holes in a cavity, we calculate the Ruppeiner invariant of a general black hole, i.e., a Kerr-Newman black hole. The Kerr-Newman metric is an exact solution of the Einstein-Maxwell equation, which describes stationary spacetime with an angular momentum and an electric charge \cite{Newman:1965my}. Later on, with the advent of the AdS/CFT conjecture \cite{Maldacena:1997re,Witten:1998qj}, the thermodynamics of a Kerr-Newman black hole in AdS space was studied in \cite{Caldarelli:1999xj}, where the Hawking-Page phase transition was found in a grand canonical ensemble, and a van de Waals-like phase transition was found in a canonical ensemble. Other thermodynamic properties of the Kerr-Newman black hole under different conditions were extensively studied \cite{Davies:1989ey,Sahay:2010wi,Cembranos:2011sr,Cheng:2016bpx,Biro:2019rms,Garcia-Compean:2020gii}. It was proposed by Smarr that under some specific conditions, a two-dimensional hypersurface of the Kerr spacetime can not be globally embedded in the three-dimensional Euclidean space  \cite{Smarr:1973zz}, therefore only the quasi-local
energy with a small angular momentum of the Kerr black hole was obtained \cite{Martinez:1994ja}. A counterterm prescription inspired by the AdS/CFT conjecture was proposed \cite{Kraus:1999di,Mann:1999pc,Balasubramanian:1999re,Lau:1999dp}, which solved this issue and was used to calculate the quasi-local energy and thermodynamic quantities of various rotating black holes \cite{Dehghani:2001af,Dehghani:2002np,Dehghani:2002nt}.

In this work, we investigate the phase transitions and thermodynamic geometry of a Kerr-Newman black hole in a cavity as well as two reduced cases, i.e., a RN black hole in a cavity and a Kerr black hole in a cavity. The quasi-local energy we adopt is defined by the Brown-York method. To make the Hamiltonian well defined, a counterterm prescription is considered. The following part of this paper is organized as follows: We first derive the quasi-local energy for a Kerr-Newman black hole and construct the first law of thermodynamics in section \ref{qet}. In section \ref{pt}, we study the phase transition in a canonical ensemble based on the quantities defined in the last section. Two reduced cases are also discussed. In section \ref{RGsect}, we probe the microscopic interactions of the black holes with the help of the thermodynamic geometry. The closing remarks are presented in section \ref{closr}.

\section{Quasi-local Energy and Thermodynamics}\label{qet}

\begin{figure}[ptb]
	\begin{center}
		\subfigure{
			\includegraphics[width=0.2\textwidth]{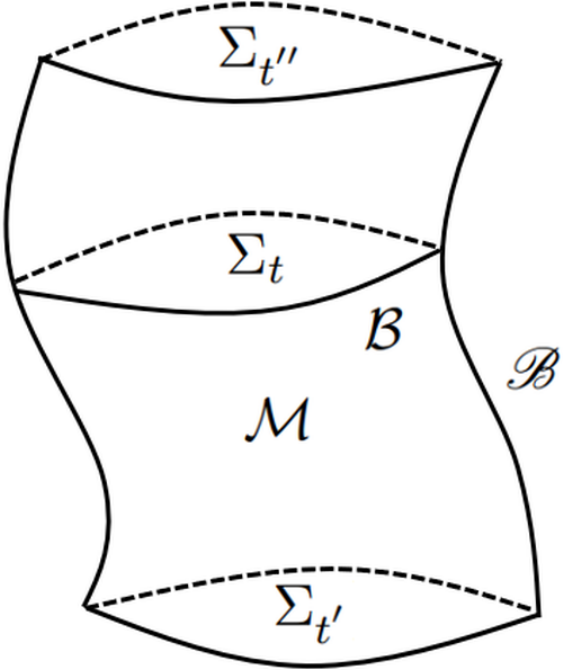}}
		\caption{Manifold $\mathcal{M}$ and its foliation. The boundary $\partial \mathcal{M}=(-\Sigma_{t^{'}})\cup\Sigma_{t^{''}}\cup \mathscr{B}$. The boundary of the spacelike hypersurface $\Sigma_{t}$ that embedded in $\mathcal{M}$ is $\mathcal{B}$.}
		\label{foliation}
	\end{center}
\end{figure}

We start from the volume gravitational action coupled to the electromagnetic field,
\begin{equation}
	\mathcal{S}_{v}=\frac{1}{16\pi}\int_{\mathcal{M}}d^4 x\sqrt{-g}\left(R-F^{\mu\nu}F_{\mu\nu}\right),
\end{equation}
where $R$ is the Ricci scalar of the manifold $\mathcal{M}$, and $F_{\mu\nu}=\partial_{\mu}A_{\nu}-\partial_{\nu}A_{\mu}$ is the electromagnetic field tensor. To give the correct field equation, a boundary term is required, and it is given by
\begin{equation}
	\mathcal{S}_{b}=\frac{1}{8\pi}\int_{\partial \mathcal{M}}d^3 x\varepsilon\sqrt{|h|}K.
\end{equation}
The integral of the above formula is calculated on the boundary $\partial \mathcal{M}$, which can be written as a union of three parts $\partial \mathcal{M}=(-\Sigma_{t^{'}})\cup\Sigma_{t^{''}}\cup \mathscr{B}$, where $\Sigma_{t^{'}}$, $\Sigma_{t^{''}}$ are two spacelike boundaries, and $\mathscr{B}$ is a timelike boundary. Besides, $K$ is the trace of the extrinsic curvature on $\partial\mathcal{M}$ as embedded in $\mathcal{M}$, and $\varepsilon=1$ if $\partial \mathcal{M}$ is timelike while $\varepsilon=-1$ if $\partial\mathcal{M}$ is spacelike. The foliation of the manifold is presented in FIG. \ref{foliation}. It turns out that the value of the action is divergent if one extends the boundary to infinity. Therefore, an extra term is needed to eliminate this divergence. Although this term is not always uniquely determined \cite{Chan:1996sx}, a natural choice is
\begin{equation}
	\mathcal{S}_0=-\frac{1}{8\pi}\int_{\partial \mathcal{M}}d^3 x\varepsilon\sqrt{|h|}K_0,\label{S0}
\end{equation}
where $K_0$ is the extrinsic curvature of $\partial\mathcal{M}$ that embedded in flat spacetime. This term makes the value of the action vanish when spacetime is flat. For a Kerr spacetime, not all the two-boundary surfaces are possible to be globally embedded in three-dimensional Euclidean space \cite{Smarr:1973zz}, which makes the extra term not well defined, and hence only the quasi-local energy with a small angular momentum is obtained \cite{Martinez:1994ja}. To address these difficulties, one can resort to a counterterm prescription, which suggests that the term $\mathcal{S}_{0}$ in Eq. (\ref{S0}) should be replaced by \cite{Kraus:1999di,Mann:1999pc,Balasubramanian:1999re}
\begin{equation}
	\mathcal{S}_{ct}=-\frac{1}{8\pi}\int_{\partial \mathcal{M}}d^3 x\varepsilon\sqrt{|h|}\sqrt{2\mathscr{R}},\label{counteraction}
\end{equation}
where $\mathscr{R}$ denotes the Ricci scalar on the three-boundary $\mathscr{B}$. Some analyses focus on another similar counterterm that substitutes $\mathscr{R}$ with the Ricci scalar of the two-boundary $\mathcal{B}$, which is also called a lightcone reference \cite{Lau:1999dp}. In the following part, we will calculate thermodynamic quantities based on the counterterm in Eq. (\ref{counteraction}). Geometric quantities in this paper are shown in TABLE \ref{gequ}.

\begin{table}[htbp]
	\scriptsize
	\begin{tabular}{ccccccc}
		\hline
		\hline
		\text{ }\text{ }\text{Manifold}\text{ }\text{ }&\text{ }\text{ }\text{Metric}\text{ }\text{ }&\text{ }\text{ }\text{Covariant Derivative}\text{ }\text{ }&\text{ }\text{ }\text{Curvature}\text{ }\text{ }&\text{ }\text{ }\text{Extrinsic Curvature}\text{ }\text{ }&\text{ }\text{ }\text{Unit Normal}\text{ }\text{ }&\text{ }\text{ }\text{Tangent Vector}\text{ }\text{ }\\
		\hline
		$\mathcal{M}$ & $g_{\alpha\beta}$ &$_{;\alpha}$  & $R$ &  & & \\
		$\Sigma_{t}$ & $h_{ab}$&$_{|a}$ & $\mathcal{R}$ & $K_{ab}$ & $n_{\alpha}$ & $e^{\alpha}_{a}$\\
		$\mathscr{B}$ & $\gamma_{ij}$ & & $\mathscr{R}$ & $\mathscr{K}_{ij}$ & $r_{\alpha}$ & $e^{\alpha}_{i}$\\
		$\mathcal{B}$ & $\sigma_{AB}$ & & $\mathscr{R}_{L}$ & $k_{AB}$ & $r_{a}$ & $e^{\alpha}_{A}$\\
		\hline
		\hline
	\end{tabular}
	\centering
	\caption{Geometric quantities in this paper.}
	\label{gequ}
\end{table}

To obtain the Hamiltonian from the action, one performs a (3 + 1) decomposition on the manifold $\mathcal{M}$ and arrives at \cite{poisson2004relativist}
\begin{equation}
	\begin{aligned}
		\mathcal{S}=\frac{1}{16\pi}\int_{t^{'}}^{t^{''}}dt\left[\int_{\Sigma_t}d^3 x \left(\mathcal{R}+K^{ab}K_{ab}-K^2-F^{\mu\nu}F_{\mu\nu}\right)N\sqrt{h}+2\int_{\mathcal{B}}d^2x\left(k-\sqrt{2\mathscr{R}}\right)N\sqrt{\sigma}\right],
	\end{aligned}
\end{equation}
where $\Sigma_{t}$ describes a constant time hypersurface, $\mathcal{R}$ is the Ricci scalar on $\Sigma_{t}$, $N$ is the lapse function, $\sigma$ denotes the determinant of the metric of $\mathcal{B}$, and $k$ is the trace of the extrinsic curvature of $\mathcal{B}$ as embedded in $\Sigma_t$. Regarding $h_{ab}$ as the dynamic variable of the gravitational field, one obtains the corresponding conjugated momentum
\begin{equation}
	p^{ab}=\frac{\partial \left(\sqrt{-g}\mathcal{L}_G\right)}{\partial \dot{h}_{ab}}=\frac{1}{16\pi}\sqrt{h}(K^{ab}-Kh^{ab}),
\end{equation}
where $\mathcal{L}_G$ is the Lagrangian of the pure gravitational field. The dot over $h_{ab}$ denotes the Lie derivative along $t^{\alpha}=Nn^{\alpha}+N^{a}e^{\alpha}_{a}$, where $N^{a}$ is the shift function. The dynamic variable of the pure electromagnetic field is $\partial_0 A_{a}$, and hence the conjugated momentum is
\begin{equation}
	\mathcal{P}^{a}=-\frac{\sqrt{-g}}{16\pi}\frac{\partial \left(F^{\mu\nu}F_{\mu\nu}\right)}{\partial \left(\partial_0 A_{a}\right)}=-\frac{\sqrt{-g}}{4\pi}F^{0a}.
\end{equation}
Consequently, the Hamiltonian density can be written as
\begin{equation}
	\mathcal{H}=p^{ab}\dot{h}_{ab}+\mathcal{P}^{a}\partial_0 A_{a}-\sqrt{-g}\left(\mathcal{L}_G-F^{\mu\nu}F_{\mu\nu}\right).
\end{equation}
Integrating the Hamiltonian density in the space region $\Sigma_t$ and adding the boundary term, we obtain the Hamiltonian of the system,
\begin{equation}
	\begin{aligned}
		H&=\frac{1}{16\pi}\int_{\Sigma_t}d^3x\sqrt{h}\left(N\mathcal{C}-2N_a\mathcal{C}^{a}+A_{0}\mathcal{G}\right)\\
		&-\frac{1}{8\pi}\int_{\mathcal{B}}d^2x\sqrt{\sigma}\left[N\left(k-\sqrt{2\mathscr{R}}\right)-N_a\left(K^{ab}-Kh^{ab}\right)r_b-\frac{8\pi}{\sqrt{h}}A_{0}r_{a}\mathcal{P}^{a}\right],
	\end{aligned}
\end{equation}
where $r_a$ is the unit normal of $\mathcal{B}$ as embedded in $\Sigma_t$. Furthermore, we have two constraint equations from general relativity and one from electromagnetism. The Hamiltonian constraint reads
\begin{equation}
	\mathcal{C}=K^{ab}K_{ab}-K^2-\mathcal{R}+16\pi\rho=0,
\end{equation}
where $\rho=T_{\alpha\beta}n^{\alpha}n^{\beta}$ is the energy density of the electromagnetic field, and $T_{\alpha\beta}$ denotes the energy-momentum tensor,
\begin{equation}
	T^{\alpha\beta}=\frac{1}{4\pi}g_{\mu\nu}F^{\alpha\mu}F^{\beta\nu}-\frac{1}{16\pi}g^{\alpha\beta}F^{\mu\nu}F_{\mu\nu}.
\end{equation}
The momentum constraint reads
\begin{equation}
	\mathcal{C}^{a}=\left(K^{ab}-Kh^{ab}\right)_{|b}-8\pi j^{a}=0,
\end{equation}
where $j_{a}=T_{\alpha\beta}e^{\alpha}_{a}n^{\beta}$ is the momentum density of the field. The Gauss's-law constraint reads
\begin{equation}
	\mathcal{G}=-\frac{16\pi}{\sqrt{h}}\mathcal{P}^{a}{}_{,a}=0.
\end{equation}
With these constraint equations, the Hamiltonian reduces to
\begin{equation}
	H=-\frac{1}{8\pi}\int_{\mathcal{B}}d^2x\sqrt{\sigma}\left[N\left(k-\sqrt{2\mathscr{R}}\right)-N_a\left(K^{ab}-Kh^{ab}\right)r_b-\frac{8\pi}{\sqrt{h}}A_{0}r_{a}\mathcal{P}^{a}\right].\label{valueH}
\end{equation}
The mass-energy of the gravitational system inside the boundary $\mathcal{B}$ is defined as the first term in the Hamiltonian (\ref{valueH}) with the choice of lapse $N=1$ \cite{Brown:1992br,Szabados:2004xxa},
\begin{equation}
	\begin{aligned}
		E&=-\frac{1}{8\pi}\int_{\mathcal{B}}d^2x\sqrt{\sigma}\left(k-\sqrt{2\mathscr{R}}\right),\label{quasenerg}
	\end{aligned}
\end{equation}
which will be regarded as the internal energy of the thermodynamic system.

We consider the Kerr-Newman spacetime, whose metric in the Boyer-Lindquist-type coordinates is given by \cite{Newman:1965my,poisson2004relativist}
\begin{equation}
	\begin{aligned}
		ds^2=-\frac{\Delta}{\rho^2}\left(dt-a\sin^2{\theta}d\phi\right)^2+&\frac{\rho^2}{\Delta}dr^2+\rho^2 d\theta^2+\frac{\sin^2{\theta}}{\rho^2}\Big[adt-\left(r^2+a^2\right)d\phi\Big]^2,\\
		A_{\alpha}dx^{\alpha}&=-\frac{Q r}{\rho^2}\left(dt-a \sin^2{\theta}d\phi\right),
	\end{aligned}
\end{equation}
with
\begin{equation}
	\begin{aligned}
		\rho^2&=r^2+a^2\cos^2{\theta},\\
		\Delta&=r^2-2Mr+a^2+Q^2,\label{Delta}
	\end{aligned}
\end{equation}
where $M$ is the ADM mass, $a=J/M$ is the magnitude of the angular momentum per unit mass, and $Q$ is the charge. The event horizon radius $r_+$ is determined by a coordinate singularity $\Delta=0$, i.e., $r_+^2-2Mr_++a^2+Q^2=0$. The induced line element on the constant time hypersurface $\Sigma_t$ can be written as
\begin{equation}
	ds^2|_{\Sigma_t}=\frac{\rho^2}{\Delta}dr^2+\rho^2 d\theta^2+\frac{\sin^2{\theta}}{\rho^2}\left[\left(r^2+a^2\right)^2-a^2 \Delta \sin^2{\theta}\right]d\phi^2.
\end{equation}
We consider that the black hole is enclosed by a cavity with a fixed radius $r_B$. The line element on the hypersurface $\mathcal{B}$, which is the wall of the cavity thus is given by
\begin{equation}
	\begin{aligned}
		ds^2|_{\mathcal{B}}=\left(r_B^2+a^2 \cos^2{\theta}\right) d\theta^2+\frac{\sin^2{\theta}}{r_B^2+a^2 \cos^2{\theta}}\left[\left(r_B^2+a^2\right)^2-a^2 \left(r_B^2-2Mr_B+a^2+Q^2\right) \sin^2{\theta}\right]d\phi^2.
	\end{aligned}
\end{equation}
The determinant of the metric of $\mathcal{B}$ can be extracted from the line element above,
\begin{equation}
	\sigma=a^4 \sin ^2{\theta}\cos^2{\theta}+2 a^2 m r_B \sin ^4{\theta}-a^2 Q^2 \sin ^4{\theta}-a^2 r_B^2 \sin ^4{\theta}+2 a^2 r_B^2 \sin ^2{\theta}+r_B^4 \sin ^2{\theta}.
\end{equation}
The trace of the extrinsic curvature $k_{ab}$ of $\mathcal{B}$ that embedded in $\Sigma_{t}$ is
\begin{equation}
	k=r^{a}{}_{|a}=\frac{a^2 \sin ^2{\theta} \left(r_B-M\right)-2 r_B \left(a^2+r_B^2\right)}{a^2 \sin ^2{\theta} \left(a^2-2 M r_B+Q^2+r_B^2\right)-\left(a^2+r_B^2\right)^2}\times  \left(\frac{a^2-2 M r_B+Q^2+r_B^2}{a^2 \cos ^2{\theta}+r_B^2}\right)^{1/2}.
\end{equation}
The Ricci scalar of the boundary $\mathscr{B}$ with a fixed radius arrives at
\begin{equation}
	\mathscr{R}=\frac{2 r_B^4+2a^2 \left(r_B^2-2 M r_B+Q^2\right) \cos^2{\theta} }{\left(r_B^2+a^2 \cos^2{\theta}\right)^3}.
\end{equation}
With these preliminaries, we then calculate the energy of the black hole in a cavity. A straightforward calculation of Eq. (\ref{quasenerg}) gives
\begin{equation}
	\begin{aligned}
		E&=\int_{0}^{\pi}d\theta \text{ }E_\theta\left(a,M,r_B,Q,\theta\right),\label{energycavity}
	\end{aligned}
\end{equation}
where
\begin{equation}
	\begin{aligned}
		E_\theta&\left(a,M,r_B,Q,\theta\right)=\left[\left(a^2+r_B^2\right)^2 \sin ^2{\theta}-a^2 \sin ^4{\theta} \left(a^2-2 M r_B+Q^2+r_B^2\right)\right]^{1/2}\\
		&\times \left\{\frac{2a^2 \cos^2{\theta}\left(r_B^2+Q^2-2M r_B\right)+2r_B^4}{\left(2a^2 \cos^2{\theta}+2r_B^2\right)^3}\right\}^{1/2}\\
		&+\frac{a^2 \sin ^4{\theta} (r_B-M)-2 r_B \left(a^2+r_B^2\right) \sin ^2{\theta}}{4 \left[\left(a^2+r_B^2\right)^2 \sin ^2{\theta}-a^2 \sin ^4{\theta} \left(a^2-2 M r_B+Q^2+r_B^2\right)\right]^{1/2}}\times \left(\frac{a^2-2 M r_B+Q^2+r_B^2}{a^2 \cos ^2{\theta}+r_B^2}\right)^{1/2}.\label{Etheta}
	\end{aligned}
\end{equation}
Moreover, extending the cavity radius to infinity, we find that the result is consistent with the ADM mass, 
\begin{equation}
	\lim_{r_B\to\infty}E=M.
\end{equation}
The entropy of the Kerr-Newman black hole equals to one quarter of the event horizon area,
\begin{equation}
	S=\frac{1}{4}A=\pi\left(r_+^2+a^2\right).\label{entropy}
\end{equation}

We consider the following three equations: the event horizon determined by $\Delta=0$ in Eq. ($\ref{Delta}$), the angular momentum $J=Ma$ and the entropy in Eq. ($\ref{entropy}$). Solving these equations, one obtains the rotating parameter $a$ and the ADM mass $M$ as functions of the entropy, the angular momentum and the charge: $a(S,J,Q)$ and $M(S,J,Q)$. After inserting functions $a(S,J,Q)$ and $M(S,J,Q)$ into $E(a,M,r_B,Q)$ in Eq. (\ref{energycavity}), we rewrite the energy as $E\left(S,J,Q,r_B\right)$. We then define the temperature as the derivative of the energy with respect to the entropy while keeping $J$, $Q$, $r_B$ fixed,
\begin{equation}
	T\left(S,J,Q,r_B\right)=\left(\frac{\partial E}{\partial S}\right)_{J,Q,r_B}=\int_{0}^{\pi} d\theta\left(\frac{\partial E_{\theta}}{\partial S}\right)_{J,Q,r_B}.\label{temperature}
\end{equation}
In the second equality, we have swapped the order of the derivative and the integral. Similarly, the angular velocity and the electric potential are defined as
\begin{equation}
	\begin{aligned}
		\Omega(S,J,Q,r_B)&=\int_{0}^{\pi} d\theta\left(\frac{\partial E_{\theta}}{\partial J}\right)_{S,Q,r_B},\\
		\Phi(S,J,Q,r_B)&=\int_{0}^{\pi} d\theta\left(\frac{\partial E_{\theta}}{\partial Q}\right)_{S,J,r_B}.
	\end{aligned}\label{thermoquanties}
\end{equation}
Therefore, the first law of thermodynamics can be written as
\begin{equation}
	dE=TdS+\Omega dJ+\Phi dQ.\label{firstlawthermodynamics}
\end{equation}
Without loss of generality, we set the cavity radius $r_B=1$ and keep it invariant in the following discussions to simplify the calculation.

\section{Phase Transition}\label{pt}

\begin{figure}[ptb]
	\begin{center}
		\subfigure{
			\includegraphics[width=0.35\textwidth]{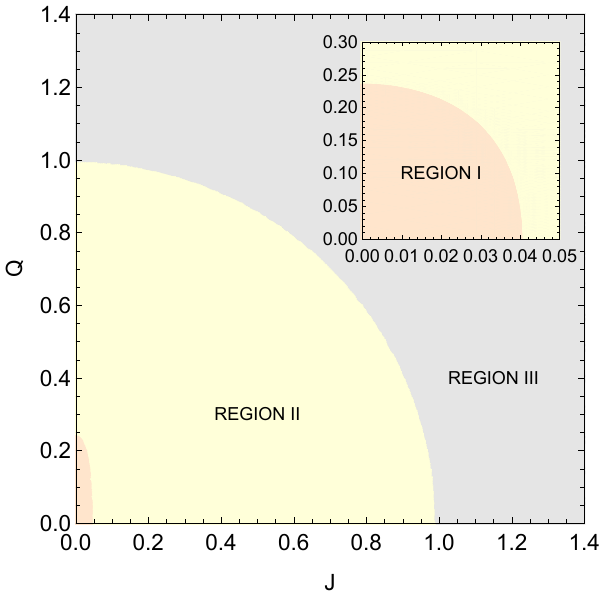}}
		\caption{Three regions in the $J$-$Q$ plane: \textbf{Region I} A first-order phase transition exists. \textbf{Region II} No phase transition but a black hole exists. \textbf{Region III} No black hole exists.}\label{JQconsplot}
	\end{center}
\end{figure}

We first analyze thermodynamic stability of the black hole in a canonical ensemble by considering the heat capacity at fixed angular momentum and the charge,
\begin{equation}
	C_{J,Q}=\lim_{\Delta T\to0}\left(\frac{\Delta Q_t}{\Delta T}\right)_{J,Q}=T\left(\frac{\partial S}{\partial T}\right)_{J,Q}=T\bigg/\left(\frac{\partial T}{\partial S}\right)_{J,Q},
\end{equation}
where $\Delta Q_t$ refers to the heat absorbed by the black hole. The sign of the heat capacity agrees with the sign of the derivative of the temperature with respect to the entropy, and so we turn to consider this term. Due to the absence of an analytic expression, we resort to numerical methods to investigate the heat capacity. We find there are three patterns for the temperature as a function of the entropy, corresponding to three regions in the $J$-$Q$ plane, as shown in FIG. \ref{JQconsplot}.

\begin{figure}[ptb]
	\begin{center}
		\subfigure{
			\includegraphics[width=0.4\textwidth]{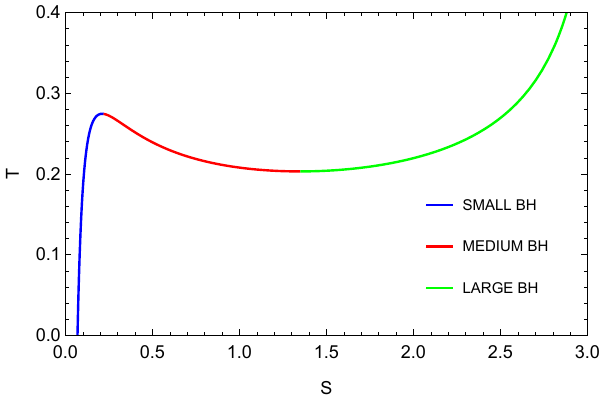}}
		\subfigure{
			\includegraphics[width=0.4\textwidth]{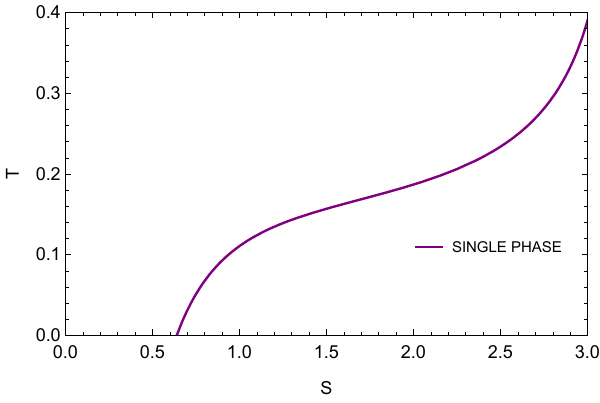}}
		\subfigure{
			\includegraphics[width=0.4\textwidth]{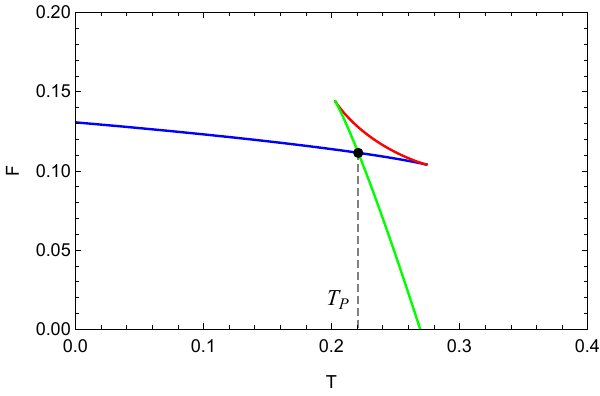}}
		\subfigure{
			\includegraphics[width=0.4\textwidth]{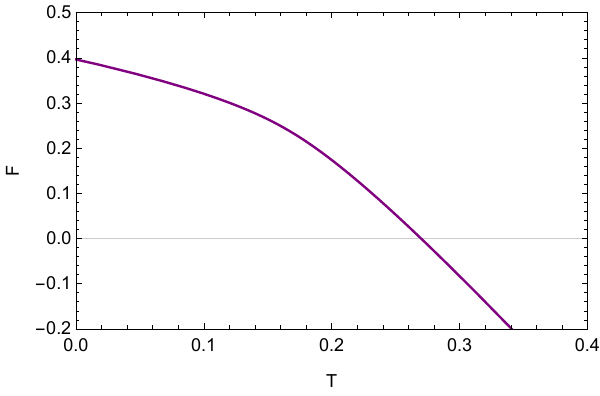}}
		\caption{Plots of the temperature $T$ with respect to the entropy $S$ (upper panels) and the free energy $F$ with respect to the temperature $T$ (lower panels) of a Kerr-Newman black hole in a cavity. \textbf{Left Column} The angular momentum $J=0.01$ and the charge $Q=0.10$ (Region I). A first-order phase transition occurs on the black dot. \textbf{Right Column} The angular momentum $J=0.10$ and the charge $Q=0.20$ (Region II).}\label{TSpteplt}
	\end{center}
\end{figure}

In Region I, the curve of the temperature with respect to the entropy can be divided into three branches, dubbed as small black hole, medium black hole and large black hole, as shown in the upper left panel of FIG. \ref{TSpteplt}. The thermodynamic stability based on the heat capacity can be read off from slopes of $T$-$S$ curves, i.e., the positive/negative slope corresponds to stable/unstable phase. Therefore, the small/medium/large black hole is thermodynamically stable/unstable/stable. In a canonical ensemble where the angular momentum $J$ and the charge $Q$ are fixed, the Helmholtz free energy, defined as $F=E-TS$ is used to investigate phase transitions. We first write the free energy as a function of $S$, $J$ and $Q$: $F(S,J,Q)$ and then plot it as a function of $T(S,J,Q)$ while regarding $S$ as the variable parameter in the lower left panel of FIG. \ref{TSpteplt}. At a low temperature, there is only a small black hole, which is thermodynamically stable. As the temperature increases, a first-order phase transition occurs from the small black hole to the large black hole. After the phase transition, the small black hole becomes a metastable phase, and the large black hole becomes the most stable phase.

In Region II, there is only one branch in the $T$-$S$ curve, as shown in the upper right panel of FIG. \ref{TSpteplt}. The branch has a positive slope, indicating a thermodynamically stable phase in this region, which is dubbed as single phase. The free energy against the temperature of the single phase is displayed in the lower right panel of FIG. \ref{TSpteplt}, which shows that there is no phase transition in Region II. In Region III, there is no black hole solution, and thus it is a non-physical region.

\begin{figure}[ptb]
	\begin{center}
		\subfigure{
			\includegraphics[width=0.4\textwidth]{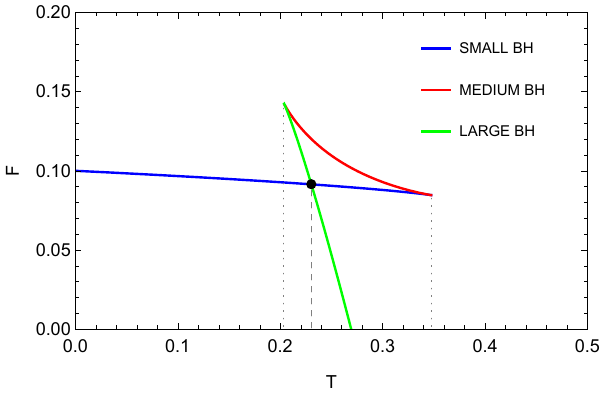}}
		\subfigure{
			\includegraphics[width=0.4\textwidth]{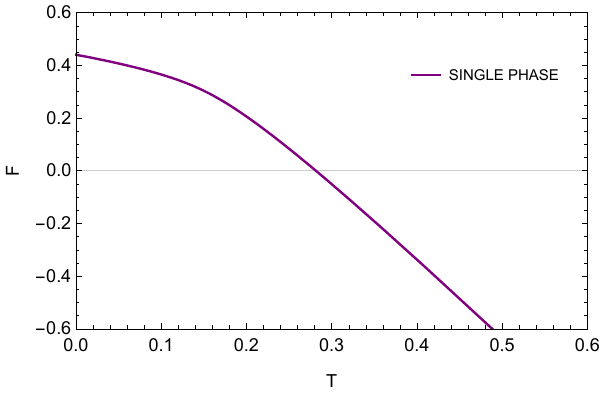}}
		\subfigure{
			\includegraphics[width=0.4\textwidth]{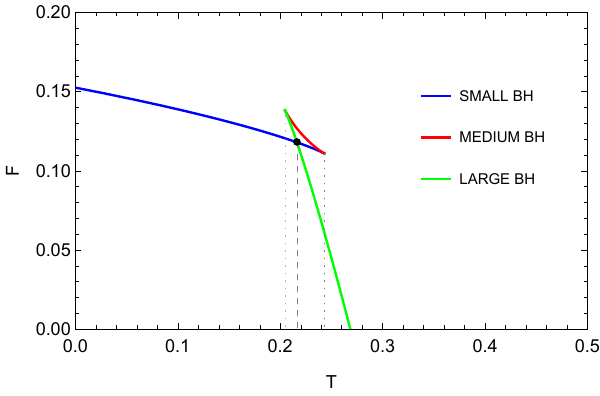}}
		\subfigure{
			\includegraphics[width=0.4\textwidth]{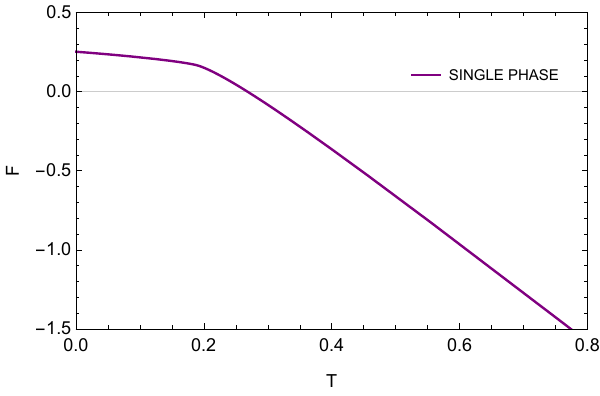}}
		\caption{\textbf{Upper Panels} Plots of the free energy $F$ with respect to the temperature $T$ of a RN black hole in a cavity. The charge $Q=0.10<Q_c$, and $Q=0.44>Q_c$ from left to right. \textbf{Lower Panels} The free energy $F$ with respect to the temperature $T$ of a Kerr black hole in a cavity. The angular momentum $J=0.02<J_c$, and $J=0.05>J_c$ from left to right. The black dot is where a first-order phase transition occurs.}\label{FTRNpte}
	\end{center}
\end{figure}

In addition, we discuss two reduced cases, i.e., a RN black hole in a cavity and a Kerr black hole in a cavity. And we will show that these black holes have similar phase structures in a cavity. Thermodynamic quantities of a RN black hole in a cavity can be obtained from those of a Kerr-Newman by making the angular momentum vanishing,
\begin{equation}
	\begin{aligned}
		E&=r_B-r_B\left(1+\frac{Q^2}{r_B^2}-\frac{\pi^{1/2}Q^2}{S^{1/2} r_B}-\frac{S^{1/2}}{\pi^{1/2} r_B}\right)^{1/2},\\
		T&=\frac{1}{4\pi^{1/2}S^{1/2}}\left(1-\frac{\pi Q^2}{S}\right)\left(1+\frac{Q^2}{r_B^2}-\frac{\pi^{1/2}Q^2}{S^{1/2} r_B}-\frac{S^{1/2}}{\pi^{1/2} r_B}\right)^{-1/2},\\
		\Phi&=\left(\frac{\pi^{1/2}Q}{S^{1/2}}-\frac{Q}{r_B}\right)\left(1+\frac{Q^2}{r_B^2}-\frac{\pi^{1/2}Q^2}{S^{1/2} r_B}-\frac{S^{1/2}}{\pi^{1/2} r_B}\right)^{-1/2},\\
		F&=r_B-r_B\left(1+\frac{Q^2}{r_B^2}-\frac{\pi^{1/2}Q^2}{S^{1/2} r_B}-\frac{S^{1/2}}{\pi^{1/2} r_B}\right)^{1/2}\\
		&-\frac{S^{1/2}}{4\pi^{1/2}}\left(1-\frac{\pi Q^2}{S}\right)\left(1+\frac{Q^2}{r_B^2}-\frac{\pi^{1/2}Q^2}{S^{1/2} r_B}-\frac{S^{1/2}}{\pi^{1/2} r_B}\right)^{-1/2}.
	\end{aligned}
\end{equation}
These results are consistent with previous studies \cite{Braden:1990hw,Wang:2020hjw}. The free energy $F$ as functions of the temperature $T$ with fixed charges can show similar swallow tail behaviors as the one of a Kerr-Newman black hole in a cavity, as shown in upper panels of FIG. \ref{FTRNpte}. When $Q<Q_c$, where $Q_c$ is the critical value of the charge, there are three different phases. The small black hole goes through a first-order phase transition and converts to the large black hole as $T$ increases. When $Q>Q_c$, there is only one single phase, and no phase transition occurs.

Similarly, thermodynamic quantities of a Kerr black hole in a cavity can be obtained by constraining the charge $Q=0$ from a Kerr-Newman black hole. These quantities are too complex to be presented here, but we have verified that they are also consistent with previous research \cite{Dehghani:2001af}. The $F$-$T$ curves of the Kerr black hole in a cavity also show similar swallowtail behavior, as shown in lower panels of FIG. \ref{FTRNpte}. When $J<J_c$, where $J_c$ is the critical value of the angular momentum, there are also three branches, corresponding three phases in the thermodynamic system. As $T$ increases, there is a small-to-large black hole phase transition. When $J>J_c$, there is only a single phase with no phase transition occurring.

\section{Thermodynamic Geometry}\label{RGsect}

In this section, we study microstructures of these black holes in a cavity with the help of the thermodynamic geometry method. Before discussing some concrete details, we first make a brief review on the thermodynamic geometry of a general thermodynamic system. The Ruppeiner metric is defined as \cite{Ruppeiner:1995zz}
\begin{equation}
	g^{R}_{\alpha\beta}=-\frac{\partial^2 S}{\partial x^{\alpha}\partial x^{\beta}},
\end{equation}
where $S$ is the entropy, and $x^{\alpha}=\left(U, V, ...\right)$ are extensive variables. The first law of thermodynamics can be written as
\begin{equation}
	dS=\frac{1}{T}dU+\frac{P}{T}dV.
\end{equation}
According to the definition of the Ruppeiner metric, the line element is
\begin{equation}
	ds^2=g_{\alpha\beta}^{R}dx^{\alpha}dx^{\beta}=\frac{1}{T}dTdS-\frac{1}{T}dPdV.
\end{equation}
The line element can be simplified to a diagonal form by choosing $S$ and $P$ as independent thermodynamic coordinates,
\begin{equation}
	\begin{aligned}
		ds^2&=\frac{1}{T}\left[\left(\frac{\partial T}{\partial S}\right)_P dS+\left(\frac{\partial T}{\partial P}\right)_S dP\right]dS-\frac{1}{T}dP\left[\left(\frac{\partial V}{\partial S}\right)_P dS+\left(\frac{\partial V}{\partial P}\right)_S dP\right]\\
		&=\frac{1}{T}\left(\frac{\partial T}{\partial S}\right)_P dS^2-\frac{1}{T}\left(\frac{\partial V}{\partial P}\right)_S dP^2,
	\end{aligned}
\end{equation}
where off-diagonal terms are eliminated by the Maxwell relation $\left(\frac{\partial T}{\partial P}\right)_S=\left(\frac{\partial V}{\partial S}\right)_P$. Having known the thermodynamic coordinates and the metric, we can construct a curvature scalar by using the definition of the Riemannian geometry, which is called the Ruppeiner invariant. A general form of the Rupppeiner invariant then reads
\begin{equation}
	R_G=\frac{T}{2\left(\frac{\partial T}{\partial S}\right)^2\left(\frac{\partial V}{\partial P}\right)}\left[\left(\frac{\partial^2 T}{\partial P^2}\right)\left(\frac{\partial^2 T}{\partial S^2}\right)-\left(\frac{\partial^2 V}{\partial S^2}\right)^2\right]-\frac{T}{2\left(\frac{\partial T}{\partial S}\right)\left(\frac{\partial V}{\partial P}\right)^2}\left[\left(\frac{\partial^2 V}{\partial S^2}\right)\left(\frac{\partial^2 V}{\partial P^2}\right)-\left(\frac{\partial^2 T}{\partial P^2}\right)^2\right],\label{Ruinv}
\end{equation}
which, apparently has been written as a function of $S$ and $P$, and this will be very helpful for our next discussions. Interestingly, the Ruppeiner invariant has an empirical conclusion: The positive/negative curvature scalar reflects the repulsive/attractive interaction of the microsystem, which implies that we can get some information about microscopic interactions of black holes by computing the Ruppeiner invariant. In the following part of this paper, we first discuss two simplified cases, that is, a RN black hole in a cavity and a Kerr black hole in a cavity. Although the thermodynamic geometry of the former case has been studied in \cite{Wang:2019cax}, we think it deserves to review this result here and compare it with the Kerr and Kerr-Newman cases.

\subsection{RN Black Hole in a Cavity}

\begin{figure}[ptb]
	\begin{center}
		\subfigure{
			\includegraphics[width=0.32\textwidth]{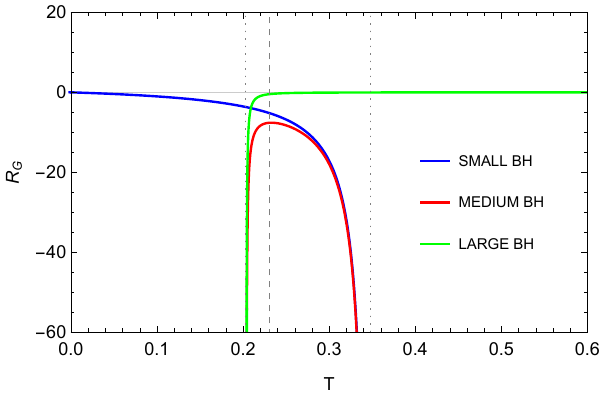}}
		\subfigure{
			\includegraphics[width=0.32\textwidth]{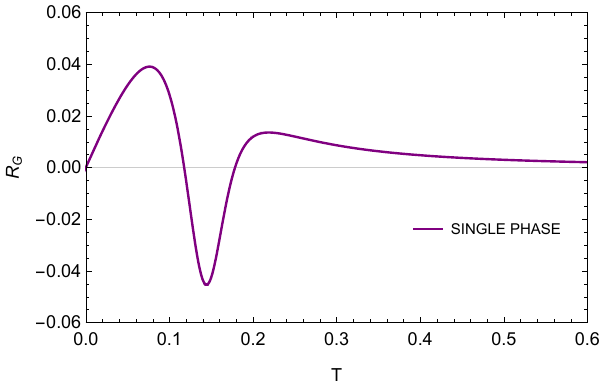}}
		\subfigure{
			\includegraphics[width=0.32\textwidth]{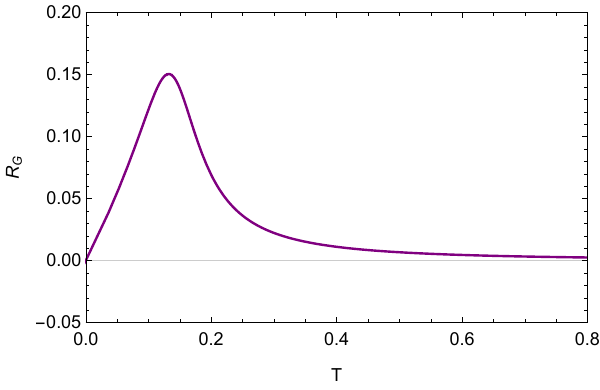}}
		\caption{Plots of the Ruppeiner invariant $R_G$ with respect to the temperature $T$ of a RN black hole in a cavity. \textbf{Left Panel} $Q=0.10<Q_c$. \textbf{Medium Panel} $Q=0.44>Q_c$. \textbf{Right Panel} $Q=0.50>Q_c$.}\label{RTRNpte}
	\end{center}
\end{figure}

The Ruppeiner invariant vanishes for a RN black hole, indicating that it might be a thermodynamic system with no interactions \cite{Aman:2003ug}. It was then argued that the charge $Q$ and potential $\Phi$ in a RN black hole in fact play the role of pressure $P$ and volume $V$ in the corresponding van der Waals fluid system, therefore a correspondence $(P,V)\rightarrow(Q,\Phi)$ was set up, and a curved thermodynamic geometry was observed for a RN black hole \cite{Shen:2005nu}. Accordingly, the internal energy should be corrected so as to satisfy the new first law of thermodynamics,
\begin{equation}
	dU=d(E-Q\Phi)=TdS-Qd\Phi.
\end{equation}
We adopt this viewpoint here, and the Ruppeiner invariant of a RN black hole in a cavity can be straightforwardly obtained by substituting $(P,V)$ into $(Q,\Phi)$ in Eq. (\ref{Ruinv}),
\begin{equation}
	\footnotesize
	R_G=\frac{\left(\pi  Q^2-S\right) \left(\pi^{1/2} r_B-S^{1/2}\right) \left(\pi^{1/2} Q^2-r_B S^{1/2}\right) \left(8 \pi  Q^2 r_B S+5 \pi ^2 Q^4 r_B-12 \pi ^{3/2} Q^4 S^{1/2}-4 \pi^{1/2} r_B^2 S^{3/2}+3 r_B S^2\right)}{S \left(2 \pi^{1/2} S^{3/2} \left(Q^2+r_B^2\right)-6 \pi ^{3/2} Q^2 S^{1/2} \left(Q^2+r_B^2\right)+6 \pi  Q^2 r_B S+5 \pi ^2 Q^4 r_B-3 r_B S^2\right)^2}.
\end{equation}
The Ruppeiner invariant $R_G$ as functions of the temperature $T$ are exhibited in FIG. \ref{RTRNpte}. When $Q<Q_c$, there are three branches in $R_G$-$T$ curves. For the small black hole phase, $R_G$ decreases with the increase of $T$ and diverges at the maximal temperature of the small black hole phase. For the medium black hole phase, there are two divergent points for $R_G$ which locate at the minimal and maximal temperature of the medium black hole. For the large black hole phase, $R_G$ diverges at the minimal temperature of the large black hole and increases as $T$ increases. When $Q>Q_c$, there is only a single phase branch in $R_G$-$T$ curves.

The phase diagram of a RN black hole in a cavity in $Q$-$T$ plane is shown in the left panel of FIG. \ref{TQregplt}. The black line describes the behavior of the temperature of a first-order phase transition against the charge. When $Q<Q_c$, the small black hole stabilizes in the region with the temperature lower than the black line while the large black hole stabilizes in the region with a temperature higher than the black line. The first-order phase transition line is a monotonically decreasing function of the charge and finally terminates at a critical point as the charge increases. It shows that the Ruppeiner invariant in the small/large black hole region is always negative, indicating the domination of an attractive interaction in the microstructure. When $Q>Q_c$, there is only the single phase in this region. For a small temperature, the Ruppeiner invariant is negative, which means that there exists an attractive interaction. As the charge increases, the Ruppeiner invariant gradually becomes positive, indicating the existence of a repulsive interaction in this region.

\subsection{Kerr Black Hole in a Cavity}

\begin{figure}[ptb]
	\begin{center}
		\subfigure{
			\includegraphics[width=0.32\textwidth]{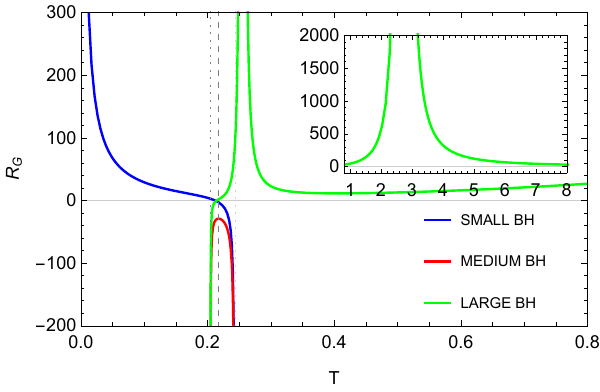}}
		\subfigure{
			\includegraphics[width=0.32\textwidth]{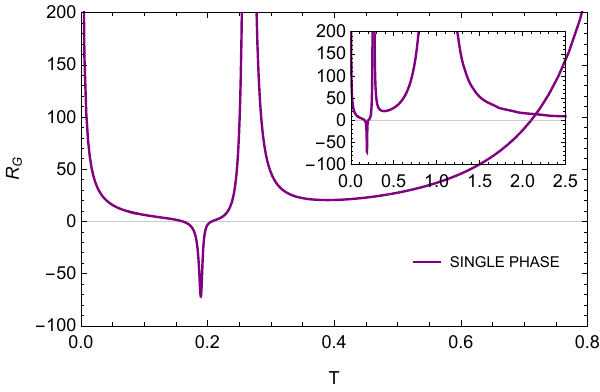}}
		\subfigure{
			\includegraphics[width=0.32\textwidth]{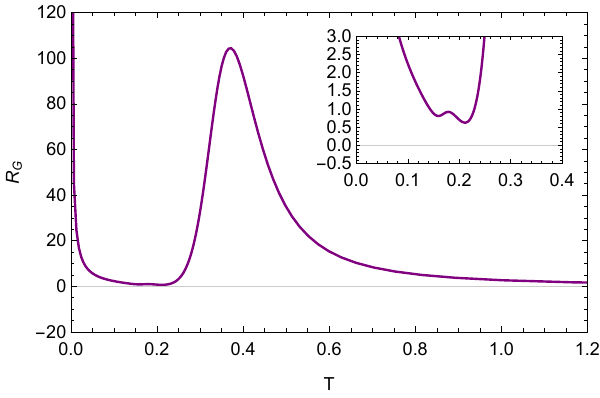}}
		\subfigure{
			\includegraphics[width=0.32\textwidth]{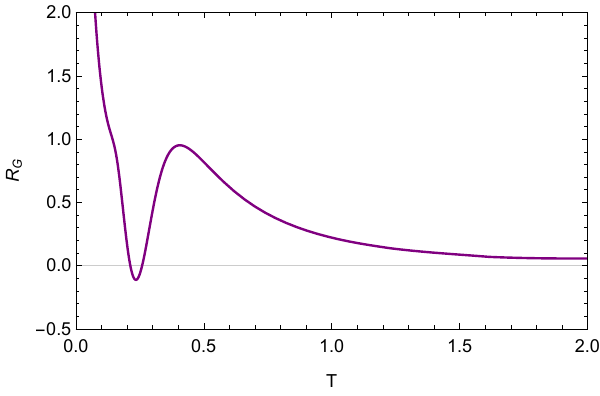}}
		\subfigure{
			\includegraphics[width=0.32\textwidth]{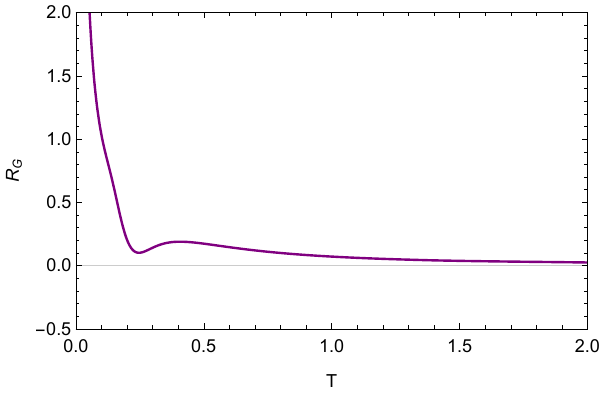}}
		\caption{Plots of the Ruppeiner invariant $R_G$ with respect to the temperature $T$ of a Kerr black hole in a cavity. \textbf{Upper Left Panel} $J=0.02<J_c$. \textbf{Upper Medium Panel} $J=0.05>J_c$. \textbf{Upper Right Panel} $J=0.10>J_c$. \textbf{Lower Left Panel} $J=0.16>J_c$. \textbf{Lower Right Panel} $J=0.23>J_c$.}\label{RTKerrpte}
	\end{center}
\end{figure}

Phase structures of a Kerr black hole in a cavity are similar to that of a RN black hole, therefore we also establish a correspondence $(P,V)\rightarrow(J,\Omega)$ as well as define the internal energy as $U=E-J\Omega$. The Ruppeiner invariant thus can be obtained from Eq. (\ref{Ruinv}) by substituting $(P,V)$ into $(J,\Omega)$. The Ruppeiner invariant $R_G$ as functions of the temperature $T$ are exhibited in FIG. \ref{RTKerrpte}. Although phase structures of a Kerr black hole and a RN black hole are very similar, the microscopic interactions show some obvious differences. A significant one is that there are some additional divergence points at some specific temperatures in some range of angular momentum. A notable one is the extreme case ($T=0$): $R_G\to\infty$ as $T\to0$ for a Kerr black hole, and $R_G\to0$ as $T\to0$ for a RN black hole.

\begin{figure}[ptb]
	\begin{center}
		\subfigure{
			\includegraphics[width=0.35\textwidth]{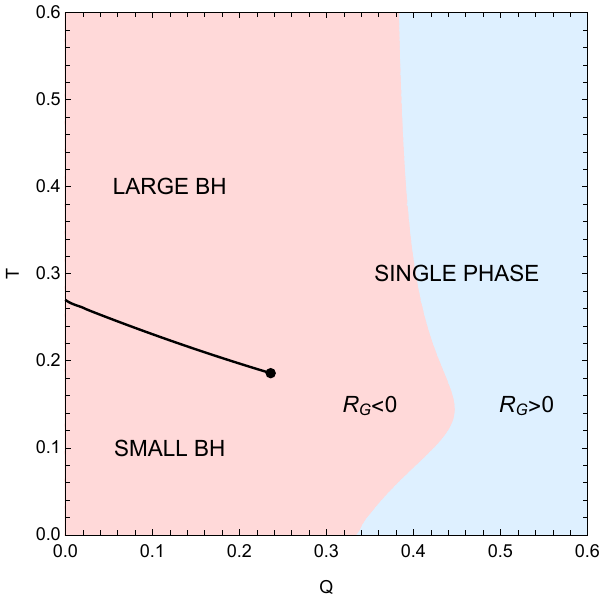}}
		\subfigure{
			\includegraphics[width=0.35\textwidth]{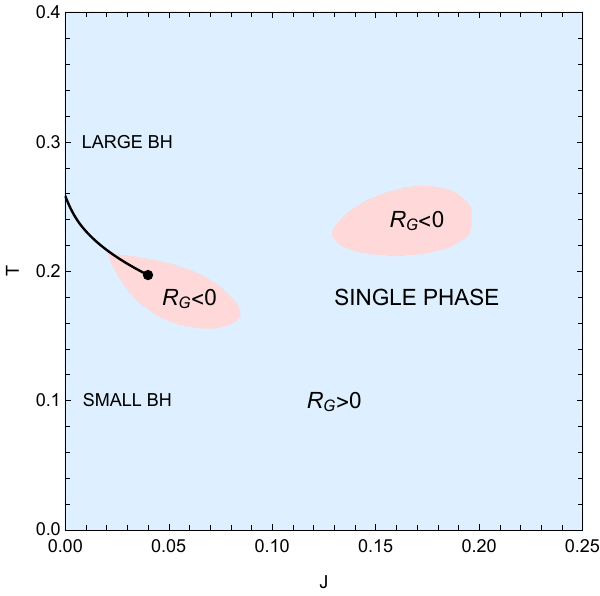}}
		\caption{\textbf{Left Panel} Phase diagram of a RN black hole in a cavity in the $Q$-$T$ plane. \textbf{Right Panel} Phase diagram of a Kerr black hole in a cavity in the $J$-$T$ plane. The black hole goes through a first-order phase transition on the black line, which terminates at a critical point as the charge or the angular momentum increases. The Ruppeiner curvature scalar $R_G$ is positive/negative in the lightblue/lightred region.}\label{TQregplt}
	\end{center}
\end{figure}

The phase diagram of a Kerr black hole in a cavity in $J$-$T$ plane is shown in the right panel of FIG. \ref{TQregplt}. As $J$ increases, the temperature of the first-order phase transition decreases. Furthermore, we surprisingly find that $R_G$ can be both positive and negative in the small/large black hole and the single phase region, which means that there are both repulsive and attractive interactions in these three phases.

\subsection{Kerr-Newman Black Hole in a Cavity}

In order to investigate the thermodynamic geometry of a Kerr-Newman black hole in a cavity and make a comparison with the RN and Kerr cases, we choose different thermodynamic coordinates so as to discuss its microstructures in different ensembles.

\subsubsection{Fixed $J$ Ensemble}

When the angular momentum $J$ is fixed, the thermodynamic coordinates are consistent with the case of a RN black hole. The internal energy is defined as $U=E-Q\Phi$. The Ruppeiner invariant can be obtained by making a substitution $(P,V)\rightarrow(Q,\Phi)$ and inserting Eq. (\ref{temperature}) and Eq. (\ref{thermoquanties}) into Eq. (\ref{Ruinv}).

We exhibit the phase diagram of a Kerr-Newman black hole in a cavity with fixed angular momentum $J=0.02$ in the left panel of FIG. \ref{TQfixJregplt}. When $J$ is fixed, the increase of $Q$ leads to the decrease of the first-order phase transition temperature. We prefer to compare this with the phase diagram of a RN black hole in a cavity since the thermodynamic coordinates are the same. Our result shows, with the introduction of an angular momentum, the microstructure changes significantly. In both small and large black hole phases, the presence of $R_G>0$ regions means that repulsive interactions can also emerge in the microstructures of these two phases.

\subsubsection{Fixed $Q$ Ensemble}

In the fixed $Q$ ensemble, the thermodynamic variables are chosen as $x^{\alpha}=(U, \Omega)$, where in the case, $U=E-\Omega J$ is defined as the internal energy of the thermodynamic system. The Ruppeiner invariant then can be obtained by substituting $(P,V)\rightarrow(J,\Omega)$ in Eq. (\ref{Ruinv}).

\begin{figure}[ptb]
	\begin{center}
		\subfigure{
			\includegraphics[width=0.35\textwidth]{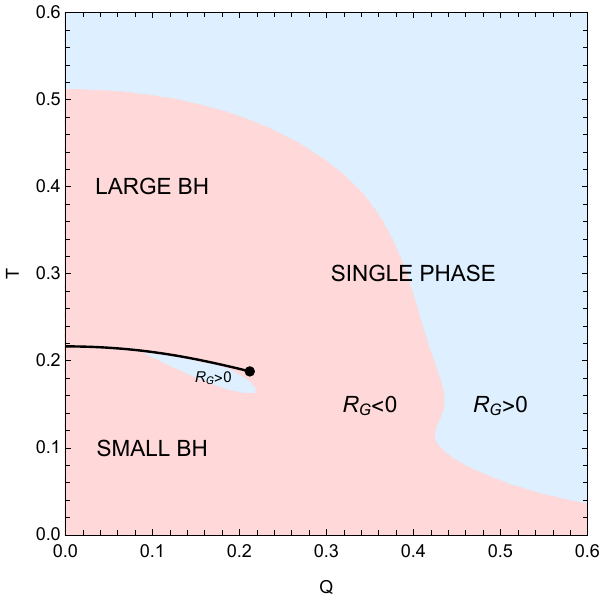}}
		\subfigure{
			\includegraphics[width=0.35\textwidth]{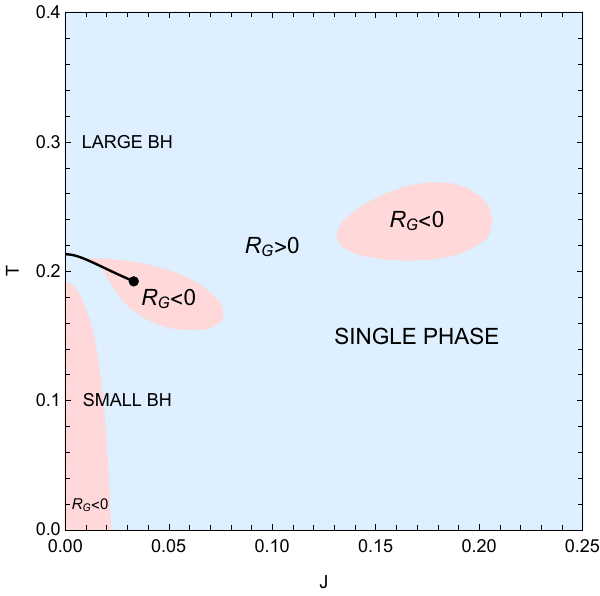}}
		\caption{\textbf{Left Panel} Phase diagram of a Kerr-Newman black hole in a cavity in the $Q$-$T$ plane with fixed angular momentum $J=0.02$. \textbf{Right Panel} Phase diagram of a Kerr-Newman black hole in a cavity in the $J$-$T$ plane with fixed charge $Q=0.15$. The black hole goes through a first-order phase transition on the black line, which terminates at a critical point as the charge or the angular momentum increases. The Ruppeiner curvature scalar $R_G$ is positive/negative in the lightblue/lightred region.}\label{TQfixJregplt}
	\end{center}
\end{figure}

The phase diagram with fixed charge $Q=0.15$ of a Kerr-Newman black hole in a cavity is shown in the right panel of FIG. \ref{TQfixJregplt}. The first-order phase transition line in $J$-$T$ plane is a monotonically decreasing function about $J$ and terminates at a critical point with the increase of $J$. Significantly different from a Kerr black hole, due to the effect of the charge, when $J$ and $T$ are small enough, there exists a region with $R_G<0$ in the small black hole phase, leads to an attractive interaction in the small black hole region.

\section{Closing Remarks}\label{closr}

In this work, we mainly studied the thermodynamic, phase transition and the thermodynamic geometry of a Kerr-Newman black hole in a cavity as well as a RN black hole in a cavity and a Kerr black hole in a cavity, and made a comparison between them. In order to derive the  quasi-local energy of a Kerr-Newman black hole in a cavity from the Hamiltonian, we resorted to a counterterm prescription, which had succeeded in defining the quasi-local energy of a Kerr spacetime in previous researches. According to the quasi-local energy, we derived thermodynamic quantities and constructed the first law of thermodynamics of the black hole.

Using the thermodynamic quantities defined above, we further analyzed the phase transition of the black hole in a cavity. It shows that there are three regions in the angular momentum-charge plane, as shown in FIG. \ref{JQconsplot}. In Region I, the black hole will go through a first-order phase transition as the temperature increases. In Region II, there is a single phase that is globally thermodynamically stable, and no phase transition occurs. In Region III, there is no black hole solution, which means that it is a non-physical region. The $T$-$S$ and $F$-$T$ curves of the black hole in Region I and Region II are shown in FIG. \ref{TSpteplt}. We concluded that the phase structure of a Kerr-Newman black hole in a cavity can show extensive similarities to the one in AdS spacetime.

The thermodynamic geometry is a helpful tool to probe microstructures of black holes. As a preparation, we first calculated the Ruppeiner invariant of a RN black hole and a Kerr black hole in a cavity. Our results show that, microstructures of the charged black hole and the rotating black hole show great differences. For a RN black hole, the microscopic interaction is dominated by an attractive interaction in the small black hole phase and the large black phase. In the single phase region, the type of the dominating interaction is determined by concrete values of the charge and the temperature, as shown in the left panel of FIG. \ref{TQregplt}. For a Kerr black hole, attractive and repulsive interactions can both exist in these three phases. When the black hole is both rotating and charged, we choosed different thermodynamic variables $x^{\alpha}$ to investigate its microstructures in different ensembles. Phase diagrams with fixed angular momentum $J$ and fixed charge $Q$ are both exhibited in FIG. \ref{TQfixJregplt}. The microstructure of a Kerr-Newman black hole in a cavity depends on thermodynamic variables $x^{\alpha}$ that we choose. When we choose $x^{\alpha}=(U,\Phi)$, the black hole exhibits microscopic interactions that are somewhat more similar to a RN black hole, and when we choose $x^{\alpha}=(U,\Omega)$, the black hole exhibits microscopic interactions more similar to a Kerr black hole. These results indicate that microstructures of black holes in a cavity are very sensitive to the selection of thermodynamic variables and their states.

\section{Acknowledgments}

We are grateful to Yihe Cao, Qi Xu and Jiayi Wu for useful discussions and valuable comments. This work is supported by NSFC (Grant No.12147207 and 12105031).

\normalem
\bibliographystyle{unsrturl}
\bibliography{Cavity}

\begin{thebibliography}{10}

\bibitem{York:1986it}
James~W. York, Jr.
\newblock {Black hole thermodynamics and the Euclidean Einstein action}.
\newblock {\em Phys. Rev. D}, 33:2092--2099, 1986.
\newblock \href {http://dx.doi.org/10.1103/PhysRevD.33.2092}
  {\path{doi:10.1103/PhysRevD.33.2092}}.

\bibitem{Hawking:1982dh}
S.~W. Hawking and Don~N. Page.
\newblock {Thermodynamics of Black Holes in anti-De Sitter Space}.
\newblock {\em Commun. Math. Phys.}, 87:577, 1983.
\newblock \href {http://dx.doi.org/10.1007/BF01208266}
  {\path{doi:10.1007/BF01208266}}.

\bibitem{Brown:1992br}
J.~David Brown and James~W. York, Jr.
\newblock {Quasilocal energy and conserved charges derived from the
  gravitational action}.
\newblock {\em Phys. Rev. D}, 47:1407--1419, 1993.
\newblock \href {http://arxiv.org/abs/gr-qc/9209012}
  {\path{arXiv:gr-qc/9209012}}, \href
  {http://dx.doi.org/10.1103/PhysRevD.47.1407}
  {\path{doi:10.1103/PhysRevD.47.1407}}.

\bibitem{Brown:1994gs}
J.~David Brown, J.~Creighton, and Robert~B. Mann.
\newblock {Temperature, energy and heat capacity of asymptotically anti-de
  Sitter black holes}.
\newblock {\em Phys. Rev. D}, 50:6394--6403, 1994.
\newblock \href {http://arxiv.org/abs/gr-qc/9405007}
  {\path{arXiv:gr-qc/9405007}}, \href
  {http://dx.doi.org/10.1103/PhysRevD.50.6394}
  {\path{doi:10.1103/PhysRevD.50.6394}}.

\bibitem{Braden:1990hw}
Harry~W. Braden, J.~David Brown, Bernard~F. Whiting, and James~W. York, Jr.
\newblock {Charged black hole in a grand canonical ensemble}.
\newblock {\em Phys. Rev. D}, 42:3376--3385, 1990.
\newblock \href {http://dx.doi.org/10.1103/PhysRevD.42.3376}
  {\path{doi:10.1103/PhysRevD.42.3376}}.

\bibitem{Carlip:2003ne}
Steven Carlip and S.~Vaidya.
\newblock {Phase transitions and critical behavior for charged black holes}.
\newblock {\em Class. Quant. Grav.}, 20:3827--3838, 2003.
\newblock \href {http://arxiv.org/abs/gr-qc/0306054}
  {\path{arXiv:gr-qc/0306054}}, \href
  {http://dx.doi.org/10.1088/0264-9381/20/16/319}
  {\path{doi:10.1088/0264-9381/20/16/319}}.

\bibitem{Lundgren:2006kt}
Andrew~P. Lundgren.
\newblock {Charged black hole in a canonical ensemble}.
\newblock {\em Phys. Rev. D}, 77:044014, 2008.
\newblock \href {http://arxiv.org/abs/gr-qc/0612119}
  {\path{arXiv:gr-qc/0612119}}, \href
  {http://dx.doi.org/10.1103/PhysRevD.77.044014}
  {\path{doi:10.1103/PhysRevD.77.044014}}.

\bibitem{Wang:2019kxp}
Peng Wang, Houwen Wu, and Haitang Yang.
\newblock {Thermodynamics and Phase Transition of a Nonlinear Electrodynamics
  Black Hole in a Cavity}.
\newblock {\em JHEP}, 07:002, 2019.
\newblock \href {http://arxiv.org/abs/1901.06216} {\path{arXiv:1901.06216}},
  \href {http://dx.doi.org/10.1007/JHEP07(2019)002}
  {\path{doi:10.1007/JHEP07(2019)002}}.

\bibitem{Liang:2019dni}
Kangkai Liang, Peng Wang, Houwen Wu, and Mingtao Yang.
\newblock {Phase structures and transitions of Born\textendash{}Infeld black
  holes in a grand canonical ensemble}.
\newblock {\em Eur. Phys. J. C}, 80(3):187, 2020.
\newblock \href {http://arxiv.org/abs/1907.00799} {\path{arXiv:1907.00799}},
  \href {http://dx.doi.org/10.1140/epjc/s10052-020-7750-z}
  {\path{doi:10.1140/epjc/s10052-020-7750-z}}.

\bibitem{Wang:2019urm}
Peng Wang, Haitang Yang, and Shuxuan Ying.
\newblock {Thermodynamics and phase transition of a Gauss-Bonnet black hole in
  a cavity}.
\newblock {\em Phys. Rev. D}, 101(6):064045, 2020.
\newblock \href {http://arxiv.org/abs/1909.01275} {\path{arXiv:1909.01275}},
  \href {http://dx.doi.org/10.1103/PhysRevD.101.064045}
  {\path{doi:10.1103/PhysRevD.101.064045}}.

\bibitem{Huang:2021iyf}
Yuchen Huang, Hongmei Jing, Jun Tao, and Feiyu Yao.
\newblock {Phase structures and transitions of quintessence surrounding RN
  black holes in a grand canonical ensemble}.
\newblock {\em Chin. Phys. C}, 45(7):075101, 2021.
\newblock \href {http://arxiv.org/abs/2104.12617} {\path{arXiv:2104.12617}},
  \href {http://dx.doi.org/10.1088/1674-1137/abf6c4}
  {\path{doi:10.1088/1674-1137/abf6c4}}.

\bibitem{Yao:2021zid}
Feiyu Yao.
\newblock {Scalarized Einstein\textendash{}Maxwell-scalar black holes in a
  cavity}.
\newblock {\em Eur. Phys. J. C}, 81(11):1009, 2021.
\newblock \href {http://arxiv.org/abs/2107.12039} {\path{arXiv:2107.12039}},
  \href {http://dx.doi.org/10.1140/epjc/s10052-021-09793-3}
  {\path{doi:10.1140/epjc/s10052-021-09793-3}}.

\bibitem{Huang:2021eby}
Yuchen Huang and Jun Tao.
\newblock {Thermodynamics and phase transition of BTZ black hole in a cavity}.
\newblock {\em Nucl. Phys. B}, 982:115881, 2022.
\newblock \href {http://arxiv.org/abs/2112.13249} {\path{arXiv:2112.13249}},
  \href {http://dx.doi.org/10.1016/j.nuclphysb.2022.115881}
  {\path{doi:10.1016/j.nuclphysb.2022.115881}}.

\bibitem{Wang:2020hjw}
Peng Wang, Houwen Wu, Haitang Yang, and Feiyu Yao.
\newblock {Extended Phase Space Thermodynamics for Black Holes in a Cavity}.
\newblock {\em JHEP}, 09:154, 2020.
\newblock \href {http://arxiv.org/abs/2006.14349} {\path{arXiv:2006.14349}},
  \href {http://dx.doi.org/10.1007/JHEP09(2020)154}
  {\path{doi:10.1007/JHEP09(2020)154}}.

\bibitem{1975JChPh..63.2479W}
F.~{Weinhold}.
\newblock {Metric geometry of equilibrium thermodynamics}.
\newblock {\em \jcp}, 63(6):2479--2483, September 1975.
\newblock \href {http://dx.doi.org/10.1063/1.431689}
  {\path{doi:10.1063/1.431689}}.

\bibitem{Ruppeiner:1995zz}
George Ruppeiner.
\newblock {Riemannian geometry in thermodynamic fluctuation theory}.
\newblock {\em Rev. Mod. Phys.}, 67:605--659, 1995.
\newblock [Erratum: Rev.Mod.Phys. 68, 313--313 (1996)].
\newblock \href {http://dx.doi.org/10.1103/RevModPhys.67.605}
  {\path{doi:10.1103/RevModPhys.67.605}}.

\bibitem{PhysRevA.20.1608}
George Ruppeiner.
\newblock Thermodynamics: A riemannian geometric model.
\newblock {\em Phys. Rev. A}, 20:1608--1613, Oct 1979.
\newblock URL: \url{https://link.aps.org/doi/10.1103/PhysRevA.20.1608}, \href
  {http://dx.doi.org/10.1103/PhysRevA.20.1608}
  {\path{doi:10.1103/PhysRevA.20.1608}}.

\bibitem{PhysRevA.41.2200}
George Ruppeiner and Christopher Davis.
\newblock Thermodynamic curvature of the multicomponent ideal gas.
\newblock {\em Phys. Rev. A}, 41:2200--2202, Feb 1990.
\newblock URL: \url{https://link.aps.org/doi/10.1103/PhysRevA.41.2200}, \href
  {http://dx.doi.org/10.1103/PhysRevA.41.2200}
  {\path{doi:10.1103/PhysRevA.41.2200}}.

\bibitem{PhysRevA.39.6515}
H.~Janyszek and R.~Mrugal/a.
\newblock Riemannian geometry and the thermodynamics of model magnetic systems.
\newblock {\em Phys. Rev. A}, 39:6515--6523, Jun 1989.
\newblock URL: \url{https://link.aps.org/doi/10.1103/PhysRevA.39.6515}, \href
  {http://dx.doi.org/10.1103/PhysRevA.39.6515}
  {\path{doi:10.1103/PhysRevA.39.6515}}.

\bibitem{1990JPhA...23..467J}
H.~{Janyszek} and R.~{Mrugaa}.
\newblock {Riemannian geometry and stability of ideal quantum gases}.
\newblock {\em Journal of Physics A Mathematical General}, 23(4):467--476,
  February 1990.
\newblock \href {http://dx.doi.org/10.1088/0305-4470/23/4/016}
  {\path{doi:10.1088/0305-4470/23/4/016}}.

\bibitem{PhysRevA.24.488}
George Ruppeiner.
\newblock Application of riemannian geometry to the thermodynamics of a simple
  fluctuating magnetic system.
\newblock {\em Phys. Rev. A}, 24:488--492, Jul 1981.
\newblock URL: \url{https://link.aps.org/doi/10.1103/PhysRevA.24.488}, \href
  {http://dx.doi.org/10.1103/PhysRevA.24.488}
  {\path{doi:10.1103/PhysRevA.24.488}}.

\bibitem{Ferrara:1997tw}
Sergio Ferrara, Gary~W. Gibbons, and Renata Kallosh.
\newblock {Black holes and critical points in moduli space}.
\newblock {\em Nucl. Phys. B}, 500:75--93, 1997.
\newblock \href {http://arxiv.org/abs/hep-th/9702103}
  {\path{arXiv:hep-th/9702103}}, \href
  {http://dx.doi.org/10.1016/S0550-3213(97)00324-6}
  {\path{doi:10.1016/S0550-3213(97)00324-6}}.

\bibitem{Cai:1998ep}
Rong-Gen Cai and Jin-Ho Cho.
\newblock {Thermodynamic curvature of the BTZ black hole}.
\newblock {\em Phys. Rev. D}, 60:067502, 1999.
\newblock \href {http://arxiv.org/abs/hep-th/9803261}
  {\path{arXiv:hep-th/9803261}}, \href
  {http://dx.doi.org/10.1103/PhysRevD.60.067502}
  {\path{doi:10.1103/PhysRevD.60.067502}}.

\bibitem{Aman:2003ug}
Jan~E. Aman, Ingemar Bengtsson, and Narit Pidokrajt.
\newblock {Geometry of black hole thermodynamics}.
\newblock {\em Gen. Rel. Grav.}, 35:1733, 2003.
\newblock \href {http://arxiv.org/abs/gr-qc/0304015}
  {\path{arXiv:gr-qc/0304015}}, \href
  {http://dx.doi.org/10.1023/A:1026058111582}
  {\path{doi:10.1023/A:1026058111582}}.

\bibitem{Shen:2005nu}
Jian-yong Shen, Rong-Gen Cai, Bin Wang, and Ru-Keng Su.
\newblock {Thermodynamic geometry and critical behavior of black holes}.
\newblock {\em Int. J. Mod. Phys. A}, 22:11--27, 2007.
\newblock \href {http://arxiv.org/abs/gr-qc/0512035}
  {\path{arXiv:gr-qc/0512035}}, \href
  {http://dx.doi.org/10.1142/S0217751X07034064}
  {\path{doi:10.1142/S0217751X07034064}}.

\bibitem{Aman:2005xk}
Jan~E. Aman and Narit Pidokrajt.
\newblock {Geometry of higher-dimensional black hole thermodynamics}.
\newblock {\em Phys. Rev. D}, 73:024017, 2006.
\newblock \href {http://arxiv.org/abs/hep-th/0510139}
  {\path{arXiv:hep-th/0510139}}, \href
  {http://dx.doi.org/10.1103/PhysRevD.73.024017}
  {\path{doi:10.1103/PhysRevD.73.024017}}.

\bibitem{Sarkar:2006tg}
Tapobrata Sarkar, Gautam Sengupta, and Bhupendra Nath~Tiwari.
\newblock {On the thermodynamic geometry of BTZ black holes}.
\newblock {\em JHEP}, 11:015, 2006.
\newblock \href {http://arxiv.org/abs/hep-th/0606084}
  {\path{arXiv:hep-th/0606084}}, \href
  {http://dx.doi.org/10.1088/1126-6708/2006/11/015}
  {\path{doi:10.1088/1126-6708/2006/11/015}}.

\bibitem{Quevedo:2008xn}
Hernando Quevedo and Alberto Sanchez.
\newblock {Geometrothermodynamics of asymptotically de Sitter black holes}.
\newblock {\em JHEP}, 09:034, 2008.
\newblock \href {http://arxiv.org/abs/0805.3003} {\path{arXiv:0805.3003}},
  \href {http://dx.doi.org/10.1088/1126-6708/2008/09/034}
  {\path{doi:10.1088/1126-6708/2008/09/034}}.

\bibitem{Ruppeiner:2008kd}
George Ruppeiner.
\newblock {Thermodynamic curvature and phase transitions in Kerr-Newman black
  holes}.
\newblock {\em Phys. Rev. D}, 78:024016, 2008.
\newblock \href {http://arxiv.org/abs/0802.1326} {\path{arXiv:0802.1326}},
  \href {http://dx.doi.org/10.1103/PhysRevD.78.024016}
  {\path{doi:10.1103/PhysRevD.78.024016}}.

\bibitem{Banerjee:2010bx}
Rabin Banerjee, Sujoy~Kumar Modak, and Saurav Samanta.
\newblock {Second Order Phase Transition and Thermodynamic Geometry in Kerr-AdS
  Black Hole}.
\newblock {\em Phys. Rev. D}, 84:064024, 2011.
\newblock \href {http://arxiv.org/abs/1005.4832} {\path{arXiv:1005.4832}},
  \href {http://dx.doi.org/10.1103/PhysRevD.84.064024}
  {\path{doi:10.1103/PhysRevD.84.064024}}.

\bibitem{Banerjee:2010da}
Rabin Banerjee, Sumit Ghosh, and Dibakar Roychowdhury.
\newblock {New type of phase transition in Reissner Nordstr\"om\textendash{}AdS
  black hole and its thermodynamic geometry}.
\newblock {\em Phys. Lett. B}, 696:156--162, 2011.
\newblock \href {http://arxiv.org/abs/1008.2644} {\path{arXiv:1008.2644}},
  \href {http://dx.doi.org/10.1016/j.physletb.2010.12.010}
  {\path{doi:10.1016/j.physletb.2010.12.010}}.

\bibitem{Niu:2011tb}
Chao Niu, Yu~Tian, and Xiao-Ning Wu.
\newblock {Critical Phenomena and Thermodynamic Geometry of RN-AdS Black
  Holes}.
\newblock {\em Phys. Rev. D}, 85:024017, 2012.
\newblock \href {http://arxiv.org/abs/1104.3066} {\path{arXiv:1104.3066}},
  \href {http://dx.doi.org/10.1103/PhysRevD.85.024017}
  {\path{doi:10.1103/PhysRevD.85.024017}}.

\bibitem{Zhang:2015ova}
Jia-Lin Zhang, Rong-Gen Cai, and Hongwei Yu.
\newblock {Phase transition and thermodynamical geometry of
  Reissner-Nordstr\"om-AdS black holes in extended phase space}.
\newblock {\em Phys. Rev. D}, 91(4):044028, 2015.
\newblock \href {http://arxiv.org/abs/1502.01428} {\path{arXiv:1502.01428}},
  \href {http://dx.doi.org/10.1103/PhysRevD.91.044028}
  {\path{doi:10.1103/PhysRevD.91.044028}}.

\bibitem{Wei:2015iwa}
Shao-Wen Wei and Yu-Xiao Liu.
\newblock {Insight into the Microscopic Structure of an AdS Black Hole from a
  Thermodynamical Phase Transition}.
\newblock {\em Phys. Rev. Lett.}, 115(11):111302, 2015.
\newblock [Erratum: Phys.Rev.Lett. 116, 169903 (2016)].
\newblock \href {http://arxiv.org/abs/1502.00386} {\path{arXiv:1502.00386}},
  \href {http://dx.doi.org/10.1103/PhysRevLett.115.111302}
  {\path{doi:10.1103/PhysRevLett.115.111302}}.

\bibitem{Wei:2019yvs}
Shao-Wen Wei, Yu-Xiao Liu, and Robert~B. Mann.
\newblock {Ruppeiner Geometry, Phase Transitions, and the Microstructure of
  Charged AdS Black Holes}.
\newblock {\em Phys. Rev. D}, 100(12):124033, 2019.
\newblock \href {http://arxiv.org/abs/1909.03887} {\path{arXiv:1909.03887}},
  \href {http://dx.doi.org/10.1103/PhysRevD.100.124033}
  {\path{doi:10.1103/PhysRevD.100.124033}}.

\bibitem{Wei:2019uqg}
Shao-Wen Wei, Yu-Xiao Liu, and Robert~B. Mann.
\newblock {Repulsive Interactions and Universal Properties of Charged
  Anti\textendash{}de Sitter Black Hole Microstructures}.
\newblock {\em Phys. Rev. Lett.}, 123(7):071103, 2019.
\newblock \href {http://arxiv.org/abs/1906.10840} {\path{arXiv:1906.10840}},
  \href {http://dx.doi.org/10.1103/PhysRevLett.123.071103}
  {\path{doi:10.1103/PhysRevLett.123.071103}}.

\bibitem{Wei:2020poh}
Shao-Wen Wei and Yu-Xiao Liu.
\newblock {Extended thermodynamics and microstructures of four-dimensional
  charged Gauss-Bonnet black hole in AdS space}.
\newblock {\em Phys. Rev. D}, 101(10):104018, 2020.
\newblock \href {http://arxiv.org/abs/2003.14275} {\path{arXiv:2003.14275}},
  \href {http://dx.doi.org/10.1103/PhysRevD.101.104018}
  {\path{doi:10.1103/PhysRevD.101.104018}}.

\bibitem{Wei:2021lmo}
Shao-Wen Wei and Yu-Xiao Liu.
\newblock {General thermodynamic geometry approach for rotating
  Kerr-anti\textendash{}de Sitter black holes}.
\newblock {\em Phys. Rev. D}, 104(8):084087, 2021.
\newblock \href {http://arxiv.org/abs/2106.06704} {\path{arXiv:2106.06704}},
  \href {http://dx.doi.org/10.1103/PhysRevD.104.084087}
  {\path{doi:10.1103/PhysRevD.104.084087}}.

\bibitem{Wang:2019cax}
Peng Wang, Houwen Wu, and Haitang Yang.
\newblock {Thermodynamic Geometry of AdS Black Holes and Black Holes in a
  Cavity}.
\newblock {\em Eur. Phys. J. C}, 80(3):216, 2020.
\newblock \href {http://arxiv.org/abs/1910.07874} {\path{arXiv:1910.07874}},
  \href {http://dx.doi.org/10.1140/epjc/s10052-020-7776-2}
  {\path{doi:10.1140/epjc/s10052-020-7776-2}}.

\bibitem{Wang:2021llu}
Peng Wang and Feiyu Yao.
\newblock {Thermodynamic geometry of black holes enclosed by a cavity in
  extended phase space}.
\newblock {\em Nucl. Phys. B}, 976:115715, 2022.
\newblock \href {http://arxiv.org/abs/2107.14640} {\path{arXiv:2107.14640}},
  \href {http://dx.doi.org/10.1016/j.nuclphysb.2022.115715}
  {\path{doi:10.1016/j.nuclphysb.2022.115715}}.

\bibitem{Newman:1965my}
E~T. Newman, R.~Couch, K.~Chinnapared, A.~Exton, A.~Prakash, and R.~Torrence.
\newblock {Metric of a Rotating, Charged Mass}.
\newblock {\em J. Math. Phys.}, 6:918--919, 1965.
\newblock \href {http://dx.doi.org/10.1063/1.1704351}
  {\path{doi:10.1063/1.1704351}}.

\bibitem{Maldacena:1997re}
Juan~Martin Maldacena.
\newblock {The Large N limit of superconformal field theories and
  supergravity}.
\newblock {\em Adv. Theor. Math. Phys.}, 2:231--252, 1998.
\newblock \href {http://arxiv.org/abs/hep-th/9711200}
  {\path{arXiv:hep-th/9711200}}, \href
  {http://dx.doi.org/10.1023/A:1026654312961}
  {\path{doi:10.1023/A:1026654312961}}.

\bibitem{Witten:1998qj}
Edward Witten.
\newblock {Anti-de Sitter space and holography}.
\newblock {\em Adv. Theor. Math. Phys.}, 2:253--291, 1998.
\newblock \href {http://arxiv.org/abs/hep-th/9802150}
  {\path{arXiv:hep-th/9802150}}, \href
  {http://dx.doi.org/10.4310/ATMP.1998.v2.n2.a2}
  {\path{doi:10.4310/ATMP.1998.v2.n2.a2}}.

\bibitem{Caldarelli:1999xj}
Marco~M. Caldarelli, Guido Cognola, and Dietmar Klemm.
\newblock {Thermodynamics of Kerr-Newman-AdS black holes and conformal field
  theories}.
\newblock {\em Class. Quant. Grav.}, 17:399--420, 2000.
\newblock \href {http://arxiv.org/abs/hep-th/9908022}
  {\path{arXiv:hep-th/9908022}}, \href
  {http://dx.doi.org/10.1088/0264-9381/17/2/310}
  {\path{doi:10.1088/0264-9381/17/2/310}}.

\bibitem{Davies:1989ey}
P.~C.~W. Davies.
\newblock {Thermodynamic Phase Transitions of {Kerr-Newman} Black Holes in De
  Sitter Space}.
\newblock {\em Class. Quant. Grav.}, 6:1909, 1989.
\newblock \href {http://dx.doi.org/10.1088/0264-9381/6/12/018}
  {\path{doi:10.1088/0264-9381/6/12/018}}.

\bibitem{Sahay:2010wi}
Anurag Sahay, Tapobrata Sarkar, and Gautam Sengupta.
\newblock {Thermodynamic Geometry and Phase Transitions in Kerr-Newman-AdS
  Black Holes}.
\newblock {\em JHEP}, 04:118, 2010.
\newblock \href {http://arxiv.org/abs/1002.2538} {\path{arXiv:1002.2538}},
  \href {http://dx.doi.org/10.1007/JHEP04(2010)118}
  {\path{doi:10.1007/JHEP04(2010)118}}.

\bibitem{Cembranos:2011sr}
J.~A.~R. Cembranos, A.~de~la Cruz-Dombriz, and P.~Jimeno~Romero.
\newblock {Kerr-Newman black holes in $f(R)$ theories}.
\newblock {\em Int. J. Geom. Meth. Mod. Phys.}, 11:1450001, 2014.
\newblock \href {http://arxiv.org/abs/1109.4519} {\path{arXiv:1109.4519}},
  \href {http://dx.doi.org/10.1142/S0219887814500017}
  {\path{doi:10.1142/S0219887814500017}}.

\bibitem{Cheng:2016bpx}
Peng Cheng, Shao-Wen Wei, and Yu-Xiao Liu.
\newblock {Critical phenomena in the extended phase space of Kerr-Newman-AdS
  black holes}.
\newblock {\em Phys. Rev. D}, 94:024025, 2016.
\newblock \href {http://arxiv.org/abs/1603.08694} {\path{arXiv:1603.08694}},
  \href {http://dx.doi.org/10.1103/PhysRevD.94.024025}
  {\path{doi:10.1103/PhysRevD.94.024025}}.

\bibitem{Biro:2019rms}
Tam\'as~S. Bir\'o, Viktor~G. Czinner, Hideo Iguchi, and P\'eter V\'an.
\newblock {Volume dependent extension of Kerr-Newman black hole
  thermodynamics}.
\newblock {\em Phys. Lett. B}, 803:135344, 2020.
\newblock \href {http://arxiv.org/abs/1912.04547} {\path{arXiv:1912.04547}},
  \href {http://dx.doi.org/10.1016/j.physletb.2020.135344}
  {\path{doi:10.1016/j.physletb.2020.135344}}.

\bibitem{Garcia-Compean:2020gii}
H.~Garc\'\i{}a-Compe\'an, V.~S. Manko, and C.~J. Ram\'\i{}rez-Valdez.
\newblock {Thermodynamics of two aligned Kerr-Newman black holes}.
\newblock {\em Phys. Rev. D}, 103(10):104001, 2021.
\newblock \href {http://arxiv.org/abs/2008.01213} {\path{arXiv:2008.01213}},
  \href {http://dx.doi.org/10.1103/PhysRevD.103.104001}
  {\path{doi:10.1103/PhysRevD.103.104001}}.

\bibitem{Smarr:1973zz}
Larry Smarr.
\newblock {Surface Geometry of Charged Rotating Black Holes}.
\newblock {\em Phys. Rev. D}, 7:289--295, 1973.
\newblock \href {http://dx.doi.org/10.1103/PhysRevD.7.289}
  {\path{doi:10.1103/PhysRevD.7.289}}.

\bibitem{Martinez:1994ja}
Erik~A. Martinez.
\newblock {Quasilocal energy for a Kerr black hole}.
\newblock {\em Phys. Rev. D}, 50:4920--4928, 1994.
\newblock \href {http://arxiv.org/abs/gr-qc/9405033}
  {\path{arXiv:gr-qc/9405033}}, \href
  {http://dx.doi.org/10.1103/PhysRevD.50.4920}
  {\path{doi:10.1103/PhysRevD.50.4920}}.

\bibitem{Kraus:1999di}
Per Kraus, Finn Larsen, and Ruud Siebelink.
\newblock {The gravitational action in asymptotically AdS and flat
  space-times}.
\newblock {\em Nucl. Phys. B}, 563:259--278, 1999.
\newblock \href {http://arxiv.org/abs/hep-th/9906127}
  {\path{arXiv:hep-th/9906127}}, \href
  {http://dx.doi.org/10.1016/S0550-3213(99)00549-0}
  {\path{doi:10.1016/S0550-3213(99)00549-0}}.

\bibitem{Mann:1999pc}
Robert~B. Mann.
\newblock {Misner string entropy}.
\newblock {\em Phys. Rev. D}, 60:104047, 1999.
\newblock \href {http://arxiv.org/abs/hep-th/9903229}
  {\path{arXiv:hep-th/9903229}}, \href
  {http://dx.doi.org/10.1103/PhysRevD.60.104047}
  {\path{doi:10.1103/PhysRevD.60.104047}}.

\bibitem{Balasubramanian:1999re}
Vijay Balasubramanian and Per Kraus.
\newblock {A Stress tensor for Anti-de Sitter gravity}.
\newblock {\em Commun. Math. Phys.}, 208:413--428, 1999.
\newblock \href {http://arxiv.org/abs/hep-th/9902121}
  {\path{arXiv:hep-th/9902121}}, \href
  {http://dx.doi.org/10.1007/s002200050764} {\path{doi:10.1007/s002200050764}}.

\bibitem{Lau:1999dp}
Stephen~R. Lau.
\newblock {Light cone reference for total gravitational energy}.
\newblock {\em Phys. Rev. D}, 60:104034, 1999.
\newblock \href {http://arxiv.org/abs/gr-qc/9903038}
  {\path{arXiv:gr-qc/9903038}}, \href
  {http://dx.doi.org/10.1103/PhysRevD.60.104034}
  {\path{doi:10.1103/PhysRevD.60.104034}}.

\bibitem{Dehghani:2001af}
M.~H. Dehghani and Robert~B. Mann.
\newblock {Quasilocal thermodynamics of Kerr and Kerr - anti-de Sitter
  space-times and the AdS / CFT correspondence}.
\newblock {\em Phys. Rev. D}, 64:044003, 2001.
\newblock \href {http://arxiv.org/abs/hep-th/0102001}
  {\path{arXiv:hep-th/0102001}}, \href
  {http://dx.doi.org/10.1103/PhysRevD.64.044003}
  {\path{doi:10.1103/PhysRevD.64.044003}}.

\bibitem{Dehghani:2002np}
M.~H. Dehghani.
\newblock {Quasilocal thermodynamics of Kerr-de Sitter space-times and the AdS
  / CFT correspondence}.
\newblock {\em Phys. Rev. D}, 65:104030, 2002.
\newblock \href {http://arxiv.org/abs/hep-th/0201128}
  {\path{arXiv:hep-th/0201128}}, \href
  {http://dx.doi.org/10.1103/PhysRevD.65.104030}
  {\path{doi:10.1103/PhysRevD.65.104030}}.

\bibitem{Dehghani:2002nt}
M.~H. Dehghani and H.~KhajehAzad.
\newblock {Thermodynamics of Kerr-Newman de Sitter black hole and dS / CFT
  correspondence}.
\newblock {\em Can. J. Phys.}, 81:1363, 2003.
\newblock \href {http://arxiv.org/abs/hep-th/0209203}
  {\path{arXiv:hep-th/0209203}}, \href {http://dx.doi.org/10.1139/p03-110}
  {\path{doi:10.1139/p03-110}}.

\bibitem{Chan:1996sx}
K.~C.~K. Chan, J.~D.~E. Creighton, and Robert~B. Mann.
\newblock {Conserved masses in GHS Einstein and string black holes and
  consistent thermodynamics}.
\newblock {\em Phys. Rev. D}, 54:3892--3899, 1996.
\newblock \href {http://arxiv.org/abs/gr-qc/9604055}
  {\path{arXiv:gr-qc/9604055}}, \href
  {http://dx.doi.org/10.1103/PhysRevD.54.3892}
  {\path{doi:10.1103/PhysRevD.54.3892}}.

\bibitem{poisson2004relativist}
Eric Poisson.
\newblock {\em A relativist's toolkit: the mathematics of black-hole
  mechanics}.
\newblock Cambridge university press, 2004.

\bibitem{Szabados:2004xxa}
Laszlo~B. Szabados.
\newblock {Quasi-Local Energy-Momentum and Angular Momentum in GR: A Review
  Article}.
\newblock {\em Living Rev. Rel.}, 7:4, 2004.
\newblock \href {http://dx.doi.org/10.12942/lrr-2004-4}
  {\path{doi:10.12942/lrr-2004-4}}.

\end{thebibliography}

\end{document}